# Simultaneous Hydrogenation and UV-photolysis Experiments of NO in CO-rich Interstellar Ice Analogues; linking HNCO, OCN$^-$, NH$_2$CHO and NH$_2$OH

G. Fedoseev[1], K.-J. Chuang[1,2], E.F. van Dishoeck[2], S. Ioppolo[3] and H. Linnartz[1]

[1]Sackler Laboratory for Astrophysics, Leiden Observatory, University of Leiden, PO Box 9513, NL-2300 RA Leiden, the Netherlands
[2]Leiden Observatory, University of Leiden, PO Box 9513, NL-2300 RA Leiden, the Netherlands
[3]Department of Physical Sciences, The Open University, Walton Hall, Milton Keynes, MK7 6AA, UK

**Abstract**
The laboratory work presented here, simulates the chemistry on icy dust grains as typical for the "CO freeze-out stage" in dark molecular clouds. It differs from previous studies in that solid-state hydrogenation and vacuum UV-photoprocessing are applied simultaneously to co-depositing molecules. In parallel, the reactions at play are described for fully characterized laboratory conditions. The focus is on the formation of molecules containing both carbon and nitrogen atoms, starting with NO in CO-, H$_2$CO-, and CH$_3$OH-rich ices at 13 K. The experiments yield three important conclusions. 1. Without UV-processing hydroxylamine (NH$_2$OH) is formed, as reported previously. 2. With UV-processing (energetic) NH$_2$ is formed through photodissociation of NH$_2$OH. This radical is key in the formation of species with an N-C bond. 3. The formation of three N-C bearing species, HNCO, OCN$^-$ and NH$_2$CHO is observed. The experiments put a clear chemical link between these species; OCN$^-$ is found to be a direct derivative of HNCO and the latter is shown to have the same precursor as formamide (NH$_2$CHO). Moreover, the addition of VUV competing channels decreases the amount of NO molecules converted into NH$_2$OH by at least one order of magnitude. Consequently, this decrease in NH$_2$OH formation yield directly influences the amount of NO molecules that can be converted into HNCO, OCN$^-$ and NH$_2$CHO.

# 1. Introduction

Isocyanic acid (HNCO) and formamide (NH$_2$CHO) have been detected in the gas-phase in a number of massive hot molecular cores (Bisschop *et al*. 2007, Ligterink *et al*. 2015a), low- and intermediate-mass pre-stellar and protostellar objects (Kahane *et al*. 2013, Lopez-Sepulcre *et al*. 2015) and in the outflow shock regions L1157-B1 and B2 (Mendoza *et al*. 2014). In these regions the gaseous molecules are thought to be directly desorbed from the ice grain mantles. Recently, both species have been searched for during the Rosetta mission on comet 67P/Churyumov–Gerasimenko. The analysis of the volatile coma composition by the ROSINA instrument yielded HNCO abundances ranging from 0.016 to 0.031 percent with respect to H$_2$O and an upper limit of 0.001 percent was derived for NH$_2$CHO (le Roy *et al*. 2015). An *in situ* analysis of the comet's surface by the COSAC instrument on board of the Philae lander reported higher values for both HNCO (0.3 percent) and NH$_2$CHO (1.8 percent) with respect to water ice (Goesmann *et al*. 2015). Although solid HNCO and NH$_2$CHO in interstellar ice have not yet been detected, the direct derivative of HNCO, the cyanate ion (OCN$^-$), has been found in the solid-state towards both low- and high-mass Young Stellar Objects (YSOs) with abundances of 0.3 to 0.8% with respect to that of H$_2$O (van Broekhuizen et al. 2004, Öberg *et al.* 2011, Boogert *et al.* 2015). This makes HNCO a strong candidate to be present in interstellar ices during the YSO formation stage. The growing astronomical interest for HNCO and NH$_2$CHO molecules is not surprising, since both molecules have direct links to astrobiology; HNCO and NH$_2$CHO are the simplest molecules that comprise of the four most abundant elements associated with living organisms, and therefore both species are often considered as possible starting points in the formation of biologically important molecules like amino acids.

Mendoza *et al.* (2014) and Lopez-Sepulcre *et al.* (2015) found a correlation between the observed HNCO and NH$_2$CHO abundances, which hints for chemical links between the two species (*i.e.*, the formation of one from the other or the formation of both from a common precursor molecule). Both studies suggested a solid state origin of NH$_2$CHO, and surface hydrogenation of HNCO was proposed as a likely source of NH$_2$CHO formation. However, recent laboratory work by Noble *et al.* (2015) contradicted HNCO hydrogenation as a direct route towards NH$_2$CHO, requiring an alternative solid state mechanism to explain the observed chemical link between NH$_2$CHO and HNCO.

Hereby we present this possible formation route with nitric oxide (NO) as a starting molecule. NO is one of the most abundant nitrogen-bearing molecules detected in dark molecular clouds and young stellar objects with abundances with respect to H$_2$ ranging from ~$10^{-9}$ to ~$10^{-7}$ (Liszt & Turner 1978, Pwa & Pottasch 1986, McGonagle *et al.* 1990, Gerin *et al.* 1992). Although less abundant than NH$_3$ and N$_2$, NO has generated much astrochemical interest in the past few years because of its higher chemical reactivity compared to other N-bearing species. The open shell configuration of an NO molecule, *i.e.*, the presence of an unpaired valence electron, substantially increases its chemical reactivity towards other species. This is particularly important under the extremely cold conditions of the interstellar medium (ISM), where gas-phase and surface reactions are strongly affected by even small endothermicity or the presence of an activation barrier. Recently, frozen NO molecules have been linked to the formation of other species, like NH$_2$OH, NO$_2$, N$_2$O, and HNO$_2$ (Joshi *et al.* 2012, Congiu *et al.* 2012a, Congiu *et al.* 2012b, Fedoseev *et al.* 2012, Minissale *et al.* 2014, Ioppolo *et al.* 2014a, Linnartz *et al.* 2015).

In space, NO molecules are primarily formed in the gas-phase through the reaction:

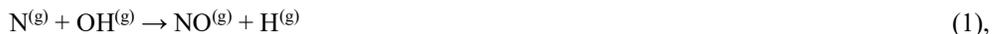

$$N^{(g)} + OH^{(g)} \rightarrow NO^{(g)} + H^{(g)} \qquad (1),$$

(see Herbst & Klemperer 1973, McGonagle *et al.* 1990, Gerin *et al.* 1992). Then, in dense regions of the ISM, NO molecules accrete onto the surface of sub-micrometer sized interstellar grains, where they in turn can react with other accreting species, like H, N, and O-atoms and/or be processed by impacting UV photons and cosmic rays (see Charnley *et al.* 2001, Congiu *et al.* 2012a). Among all the possible solid state reactions involving NO and the aforementioned atoms, those with hydrogen are the most important ones due to the great overabundance of hydrogen in comparison to other atomic species. Thus, hydrogenation of NO has been the subject of recent theoretical (Charnley *et al.* 2001, Blagojevich *et al.* 2003) and experimental (Congiu *et al.* 2012a, Congiu *et al.* 2012b, Fedoseev *et al.* 2012) studies, and was shown to result in the formation of hydroxylamine (NH$_2$OH) through three sequential steps:

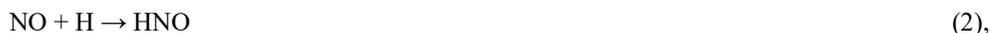

$$NO + H \rightarrow HNO \qquad (2),$$

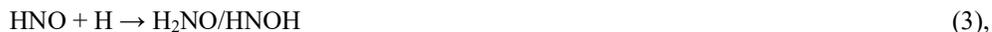

$$HNO + H \rightarrow H_2NO/HNOH \qquad (3),$$

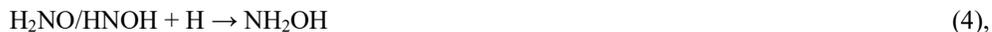

$$H_2NO/HNOH + H \rightarrow NH_2OH \qquad (4),$$

that all proceed with either no or very low activation barriers. Very recently also another formation route of hydroxylamine through O-atom addition to ammonia was discussed (He *et al.* 2015), but despite these laboratory findings dedicated astronomical searches for $NH_2OH$ towards the molecular outflow L1157 (McGuire et al. 2015) did only result in upper limits, so far. A possible explanation is that hydroxylamine is easily photodissociated and that the resulting photofragments like $NH_2$ are rapidly consumed in reactions with other species present in the ice. According to this picture, $NH_2OH$ is therefore a candidate precursor of more complex organic species, also including N-C bond bearing species. One of the aims of this work is to experimentally verify this hypothesis.

In Congiu et al. (2012b) and Fedoseev et al. (2012), hydrogenation of NO has been investigated both in the sub-monolayer and multi-layer regime, respectively. In the first case, pure NO-molecules were deposited on amorphous silicates with the aim to simulate the hydrogenation of NO on the surface of bare interstellar grains (Congiu et al. 2012b). In the second case, the simultaneous accretion of NO and $H_2O$ or CO molecules (Fedoseev et al. 2012) aims at reproducing the polar (water-rich) and apolar (water-poor) layers of ice mantles growing at different stages of dark molecular cloud evolution (Boogert et al. 2015). The CO-rich apolar layer is of greater astrochemical interest because according to current models the formation of NO in the gas phase through reaction (1) peaks when a significant amount of water-rich ice is already formed on the surface of the grains, and the abundance of OH in the gas-phase increases (Charnley et al. 2001, Visser et al. 2011, Congiu et al. 2012b, Vasyunina et al. 2012). As a consequence, the accretion of NO on the ice surface should correspond to the freeze-out of CO. Recently, Fedoseev et al. (2012) studied the possible interaction between the NO-hydrogenation reaction network and CO-molecules that can potentially be followed by the formation of N-C bearing species. However, in this laboratory study, no sign of N-C bond formation was found upon H-atom bombardment or UV-photolysis of NO in a CO-rich environment.

A possible explanation for the lack of C-N bond formation in the aforementioned experiments is the inability to study atom bombardment and UV-photolysis during a single experiment. A recent modification of our SURFRESIDE$^2$ ultra-high vacuum setup allows now investigation of both processes, sequentially and simultaneously in the same ice (Ioppolo et al. 2013). This is made possible by the addition of a $H_2$-flow microwave discharge UV-lamp to the main chamber in addition to the two existing atom beam lines. Furthermore, two independent molecular deposition lines allow for the simultaneous accretion of a selected molecule (NO) with another species, simulating a more realistic interstellar analogue composition of the bulk of the ice. The details are described in section 2.

This work presents the first simultaneous hydrogenation and UV-photolysis experiments of accreting NO molecules in three distinct molecular environments: CO-, $H_2CO$-, and $CH_3OH$-rich ices. The choice of CO is motivated by the fact that, as previously mentioned, the peak of NO accretion onto grains is expected to overlap with the accretion peak of CO (Charnley et al. 2001, Congiu et al. 2012a, Vasyunina et al. 2012). $H_2CO$ and $CH_3OH$ are the most abundant products of CO hydrogenation and therefore are expected to be present in the same ice layer according to a large number of experimental, theoretical, and observational studies (Hiraoka et al. 1994, Zhitnikov et al. 2002, Watanabe & Kouchi 2002, Fuchs et al. 2009, Cuppen et al. 2011, Penteado et al. 2015, Boogert et al. 2015, Linnartz et al. 2015). It should be noted that another ice constituent abundantly observed in CO-rich interstellar ice mantles is $CO_2$ that can be formed via the CO+OH pathway (Oba et al. 2010, Ioppolo et al. 2011, Noble et al. 2011). However, because of the high chemical inertness of $CO_2$ in comparison to CO, $H_2CO$, and $CH_3OH$ molecules, we do not include carbon dioxide in our experiments. Moreover, special care is taken to control the ratio between the used H-atom and UV-photon fluxes in an attempt to get as close as possible to the mean ratios expected for the life-time of dark molecular clouds (Prasad & Tarafdar 1983, Mathis et al. 1983, Duley & Williams 1984, Goldsmith & Li 2005, Fuchs et al. 2009). It should be noted, though, that the main focus of the present study is on the formation of various N-C bearing molecules and the photochemistry of NO hydrogenation products and not on the photochemistry of CO, $H_2CO$, and $CH_3OH$-ices that leads to the formation of carbon-bearing complex organic molecules (Öberg et al. 2011, Henderson & Gudipati 2015, Maity et al. 2015, Paardekooper et al. 2016).

The next section describes the extended setup and the experimental procedures. The results section summarizes our findings and is followed by the discussion section that focuses on the possible reaction mechanisms responsible for the formation of each of the detected N-C bearing molecules. The paper concludes with the astrochemical implications and the astrobiological importance of the newly formed molecules.

## 2. Experimental procedure

### 2.1 Experimental setup

All experiments are performed using a modified version of the SURFace Reaction Simulation Device (SURFRESIDE$^2$) UHV setup (Ioppolo *et al.* 2013, Linnartz *et al.* 2015). The original design has been described in detail by Ioppolo *et al.* (2013). The setup consists of three distinct UHV chambers separated by shutters. A gold-coated substrate is mounted on top of the cold head of a He close-cycle cryostat placed in the center of the main UHV chamber (base pressure ~$10^{-10}$ mbar). The ices are grown on the substrate with monolayer (ML) precision (where 1 ML is assumed to be $1 \times 10^{15}$ mol cm$^{-2}$). The temperature control is realized be means of heating wires. The absolute temperature precision is <2 K while the relative precision between experiments is as low as 0.5 K. Two other UHV chambers ($10^{-9}$-$10^{-10}$ mbar) contain two atom-beam sources. Two different atom sources are used to produce a number of atoms and radicals (*i.e.*, H, D, N, O, OH, NH, and NH$_2$). These are a Hydrogen Atom Beam Source (HABS, Dr. Eberl MBE-Komponenten GmbH, see Tschersich 2000) that generates atoms by cracking the parent molecules that pass through the hot (1850°C) tungsten capillary, and a Microwave Atom Source (MWAS, Oxford Scientific Ltd, see Anton *et al.* 2000) that generates atoms and radicals by using a capacitively-coupled microwave discharge (125-300 W at 2.45 GHz). The sample surface is exposed to two atom-beams with controllable fluxes and incidence angles of 45º and 135°. A nose-shaped quartz pipe is placed along the path of each of the atom-beams to quench excited electronic and ro-vibrational states of atoms and non-dissociated molecules through collisions with its walls before they reach the sample. Details on the H-atom flux calibration can be found in Ioppolo *et al.* (2013).

In addition to the two atom-beam lines, two separate molecular dosing lines are available for the deposition of stable species. These lines can be operated independently and are used to simulate different molecular environments corresponding to different stages during the interstellar ice evolution. In this work, NO (Linde 2.5), CO (Linde 2.0), and H$_2$ (Praxair 5.0) are used as gas supplies during our experiments. H$_2$CO is obtained by thermal decomposition of paraformaldehyde (Sigma-Aldrich, 95%) at 80°C under vacuum. Liquid CH$_3$OH sample (Sigma-Aldrich, 99.8%) is frozen-pumped-thawed three times before it is used as a mixture component. Gas mixtures are prepared in the pre-pumped (<$10^{-4}$ mbar) full-metal dosing line by introducing NO and a second gas component into the two distinct parts of the dosing line with a known volume ratio and by letting gases mix with each other over time. A high-precision full metal leak valve is used to introduce this gas mixture into the main chamber. The molecular masses of NO and CO (30 and 28 amu) and, therefore, the corresponding mean thermal velocities of both molecules are in close proximity. The desorption temperatures of both molecules are also close. Thus, we assume the ratio between NO and CO fluxes at the surface of the sample to be nearly equal to the NO:CO ratio in the dosing line. Although mean thermal velocities of H$_2$CO (30 amu) and CH$_3$OH (32 amu) are also similar to that of NO, the desorption temperatures of H$_2$CO and CH$_3$OH are considerably higher than that of nitric oxide and well above the temperature of the thermal shield around the sample. This will lead to a decrease of the actual H$_2$CO and CH$_3$OH deposition rates because some H$_2$CO and CH$_3$OH can deposit onto the shield as well, resulting into a somewhat higher effective NO:H$_2$CO and NO:CH$_3$OH deposition ratios, i.e., a higher fraction of NO in the resulting ice mixture compared to that in the dosing line.

Deposition proceeds under 68º and 158º incidence angles, different from the original SURFRESIDE$^2$ design where the second line was mounted at 90º. The cleared port normal to the substrate surface is now used to expose the sample to a UV-photon beam. UV photons are generated by a broadband hydrogen microwave discharge (0.5 mbar, 100 W) that peak at 121 nm (Ly-α) and 160 nm (Ligterink *et al.* 2015b). The UV-lamp is connected and separated from the UHV chamber through a $MgF_2$ window. The divergence of the beam guarantees that the whole area of the gold substrate (2.5 x 2.5 cm) is exposed to UV photons. This results in a configuration in which ices grown through two independent deposition lines, can be bombarded by radical species from two atom beam lines and can be irradiated by UV photons, sequentially or simultaneously.

The method used to evaluate the UV-photon flux at the substrate is based on a simple extrapolation of the UV-photon flux measurements performed by Ligterink et al. (2015) using a National Institute of Standards and Technology (NIST) calibrated photodiode (AXUV-100) for the same lamp configuration, microwave cavity and power supply. Assuming the UV-lamp to be a point source and taking into account the difference in distance between UV-source and ice target for SURFRESIDE$^2$ and the setup used in Ligterink et al. (2015b), a lower limit of ~$1.3 \times 10^{13}$ photons cm$^{-2}$ s$^{-1}$ is obtained. This value is used to derive exposure ratios and total UV-photon fluences summarized in Table **1**.

**2.2. Experimental methods**

Two different experimental procedures are used to study the UV-induced photochemistry of the NO hydrogenation surface reaction products: sequential UV (seq-exposure, *i.e.*, after completion of the hydrogenation) and simultaneous UV exposure (co-exposure, *i.e.*, during hydrogenation) of the ice. Furthermore, all the experiments are divided in two main sets that work with either pure NO or NO mixed with CO, $H_2CO$, and $CH_3OH$. All the relevant experiments are summarized in Table 1.

The sequential UV exposure experiments are performed in three steps. First, a 10 L (Langmuir) layer of Ar ice is deposited on top of a bare gold sample. This is done to avoid possible interactions on the interphase between the gold surface and the ice molecules upon UV exposure. Second, a co-deposition of NO molecules with an overabundance of H atoms is performed at 13 K. Alternatively, to reproduce a CO-rich interstellar ice analogue, a mixture of NO and CO is co-deposited with H atoms to form a layer of CO ice containing isolated $NH_2OH$ as well as non-fully hydrogenated HNO and NO molecules. To better compare the experimental results with each other, the same exposure dose of NO equal to 8.5 L is set for all the experiments and a total H-atom fluence equal to $6 \times 10^{16}$ atoms cm$^{-2}$ is used. Co-deposition of NO with CO, $H_2CO$, and $CH_3OH$ molecules is performed by introducing a pre-made gas-mixture in a 1:12 ratio through a single molecular dosing line. This 1:12 ratio is used to guarantee that each depositing NO molecule will have at least 12 surrounding CO molecules. This corresponds to the number of neighbors for each of the molecules in a pure CO crystal (Kohin 1960). The same 1:12 ratio is then applied to $H_2CO$ and $CH_3OH$ molecules. Furthermore the NO:CO ratio of 1:12 is close to that expected astronomically on a grain surface based on astrochemical model predictions; see for example the surface abundances of NO and CO determined for $1 \times 10^3$-$1 \times 10^5$ years in Charnley et al. (2001), and Congiu et al. (2012a). During the third and final step of a sequential UV exposure experiment, the resulting ice is exposed to the UV light of the lamp. A Reflection Absorption InfraRed (RAIR) spectrum in the range from 700 to 4000 cm$^{-1}$ is recorded with 1 cm$^{-1}$ resolution after deposition of the Ar ice layer. This spectrum is used as a reference for a series of RAIR spectra acquired every 5 minutes during the full time of the NO (or NO:CO) + H co-deposition. Once this step is completed, another reference spectrum is obtained and a second series of RAIR spectra is acquired every 2 minutes for the full time of the UV exposure.

The experiment is completed by performing a temperature programmed desorption (TPD) measurement by means of a Quadrupole Mass Spectrometer (QMS) of the photo-processed ice as a complementary analytical tool. The following procedure is systematically performed. First, an ice is desorbed with a constant rate of 2 K/min up to 50 K to gently remove volatile species from the ice. Then a 5 K/min rate is used until 225 K is reached to provide a better peak-to-noise ratio in the QMS signal. Every 6 seconds, a mass scan is taken. Simultaneously, RAIR spectra are acquired every 2 minutes for the full time of the TPD. It should be noted that the simultaneous recording of RAIRS and TPD spectra offers a highly effective tool to unambiguously identify molecules formed in the ice (see

Öberg et al. 2009 and Ioppolo et al. 2014b)

The UV co-exposure experiment consists of two steps. The first step is the same as for the sequential UV exposure experiments (*i.e.*, a deposition of 10 L of Ar ice on the bare substrate). During the second step, a simultaneous co-exposure of H atoms and UV photons with 8.5 L of NO molecules is performed. In the case of CO-, $H_2CO$-, and $CH_3OH$-rich interstellar ice analogues, mixtures of NO:CO (NO:$H_2CO$ or NO:$CH_3OH$) are co-deposited with H atoms and UV photons. Similar to the sequential UV exposure procedure, a reference RAIR spectrum is obtained after Ar deposition and a series of RAIR spectra is acquired every 5 minutes for the full time of co-exposure with UV photons. The experiments are completed by performing a QMS-TPD in the very same way as for sequential UV exposure experiments. The main aim of the UV co-exposure experiments is to study the effect of UV photons on all the NO-hydrogenation reaction products, including intermediates. These experiments are actually closer to the conditions observed in the ISM, where accreting NO molecules are simultaneously exposed to H atoms and UV photons.

A number of control experiments are performed to verify that the formed products are the result of UV-induced photochemistry (see Table 1). With this purpose in mind, all the experiments are repeated under the very same experimental conditions, but without UV-exposure. In this way, the obtained RAIR and QMS-TPD spectra can be used as a reference for comparison and the assignment of photochemistry products can be further constrained.

**Table 1.** List of discussed experiments.

| Ref. N | Experiment | UV-exposure | Ratio[a] | $T_{sample}$ (K) | $NO_{dep}$ (L s$^{-1}$) | 2nd Molecule$_{dep}$ (L s$^{-1}$) | H-atom flux ($10^{13}$ cm$^{-2}$ s$^{-1}$) | UV-flux ($10^{13}$ cm$^{-2}$ s$^{-1}$) | Time (s) | TPD |
|---|---|---|---|---|---|---|---|---|---|---|
| 1.1 | NO | - | - | 13 | 0.0017 | - | | | 5000 | QMS/RAIR |
| 1.2 | NO:H | - | 1:7 | 13 | 0.0017 | - | 1.2 | | 5000 | QMS/RAIR |
| 1.3 | NO:H + UV | seq-exposure | 1:7 | 13 | 0.0017 | - | 1.2 | 1.3 | 5000+1200 | QMS/RAIR |
| 1.4 | NO:H:UV | co-exposure | 1:7:8 | 13 | 0.0017 | - | 1.2 | 1.3 | 5000 | QMS/RAIR |
| | | | | | | $CO_{dep}$ | | | | |
| 2.1 | NO:CO:H | - | 1:12:7 | 13 | 0.0017 | 0.02 | 1.2 | | 5000 | QMS/RAIR |
| 2.2 | NO:CO:H+UV | seq-exposure | 1:12:7 | 13 | 0.0017 | 0.02 | 1.2 | 1.3 | 5000+1200 | QMS/RAIR |
| 2.3 | NO:CO:H:UV | co-exposure | 1:12:7:8 | 13 | 0.0017 | 0.02 | 1.2 | 1.3 | 5000 | QMS/RAIR |
| | | | | | | $H_2CO_{dep}$ | | | | |
| 3.1 | NO:$H_2CO$ | - | 1:12 | 13 | 0.0017 | 0.02 | | | 5000 | |
| 3.2 | NO:$H_2CO$:H | - | 1:12:7 | 13 | 0.0017 | 0.02 | 1.2 | | 5000 | QMS/RAIR |
| 3.3 | NO:$H_2CO$:H:UV | co-exposure | 1:12:7:8 | 13 | 0.0017 | 0.02 | 1.2 | 1.3 | 5000 | QMS/RAIR |
| | | | | | | $CH_3OH_{dep}$ | | | | |
| 4.1 | NO:$CH_3OH$ | - | 1:12 | 13 | 0.0017 | 0.02 | | | 5000 | |
| 4.2 | NO:$CH_3OH$:H | - | 1:12:7 | 13 | 0.0017 | 0.02 | 1.2 | | 5000 | QMS/RAIR |
| 4.3 | NO:$CH_3OH$:H:UV | co-exposure | 1:12:7:8 | 13 | 0.0017 | 0.02 | 1.2 | 1.3 | 5000 | QMS/RAIR |

[a]under the assumption that 1 L (Langmuir) exposure leads to the surface coverage of 1 ML ≈ $10^{15}$ particles cm$^{-2}$

## 3. Results

### 3.1 Simultaneous hydrogenation and UV-processing of pure NO

A set of experiments (exp. 1.1-1.4 in Table 1) is performed to investigate the effect of UV-photolysis on the non-energetic NO hydrogenation reaction route. Figure 1 compares the RAIR spectrum obtained after NO co-deposition with H atoms (Fig. 1a), the RAIR spectrum obtained after NO co-deposition with H atoms followed by UV-photon sequential exposure (Fig. 1b), and the RAIR spectrum obtained after simultaneous NO co-deposition with H atoms and co-exposure by UV photons (Fig. 1c). The RAIR spectrum obtained after co-deposition of pure NO with H atoms is similar to the spectra obtained in previous studies of surface NO hydrogenation reactions (*e.g.*, Congiu *et al.* 2012a, Fedoseev *et al.* 2012). The main product of NO hydrogenation is hydroxylamine, $NH_2OH$, a potential glycine precursor (Blagojevich *et al.* 2003, Congiu *et al.* 2012a). The formation of $NH_2OH$ is confirmed in our experiments through the presence of wide absorption bands in the range of the O-H and N-H stretching vibrations (2600 - 3400 cm$^{-1}$) and mostly through the more isolated and sharper absorption bands in the bending vibrational mode range (1700 - 900 cm$^{-1}$). These are the peaks centered around 1607, 1514, 1196, 914, and possibly 990 cm$^{-1}$ (see figure 1 of Nightingale & Wagner 1954 and figure 1 of Congiu *et al.* 2012a for comparison). Another species whose absorption features are observed in Fig. 1a is $N_2O$ (2235 and 1283 cm$^{-1}$). The presence of $N_2O$ can be explained at least partially by intrinsic contaminations in the NO gas cylinder. No signs of monomeric NO and only traces of *cis*-(NO)$_2$ dimer are observed around 1865 and 1761 cm$^{-1}$ (see Congiu *et al.* 2012a, Fedoseev *et al.* 2012).

The RAIR spectra obtained after UV-processing of the ice (Fig. 1b) clearly show that a visible fraction of the solid $NH_2OH$ is consumed and that new species form. These are $NH_3$, which can be tentatively assigned through its two absorption bands around 1641 and 1112 cm$^{-1}$ (Reding & Hornig 1955, Bertie & Shehata 1985, and Fig. 11.14 of Öberg 2009), and $H_2O$, confirmed by the detection of its bending mode centered around 1680 cm$^{-1}$ (see zoom-in panel in Fig. 1b), its libration mode (broad band below 980 cm$^{-1}$), and by a shift in the wide absorption band between 2600 - 3400 cm$^{-1}$ towards higher wavenumbers. The latter is due to the disappearance of a weaker N-H stretching vibration mode of $NH_2OH$ and the consequent appearance of a stronger O-H stretching vibration mode of $H_2O$. The assignment of $NH_3$ and $H_2O$ is further constrained by the comparison of our RAIR data (Fig. 1b) with a spectrum of a $NH_3:H_2O$ 1:4 mixture in the 1800-1300 cm$^{-1}$ spectral range taken from Öberg 2009 (Fig. 1d). Another species that is abundantly formed upon UV-processing, but cannot be detected by means of RAIRS, is $N_2$. Its presence in irradiated ice samples is confirmed by means of TPD; the QMS shows a broad $N_2$ desorption feature that starts at 29 K and peaks around 40 K (Hiraoka et al. 1995, see also Fig. 2 for more details on TPD-QMS data). Along with traces of *cis*-(NO)$_2$, also weak signals due to NO monomers can be resolved after UV-processing of the ice through their characteristic absorption at 1875 cm$^{-1}$ (Fateley *et al.* 1959, Fedoseev *et al.* 2012, Minissale *et al.* 2014).

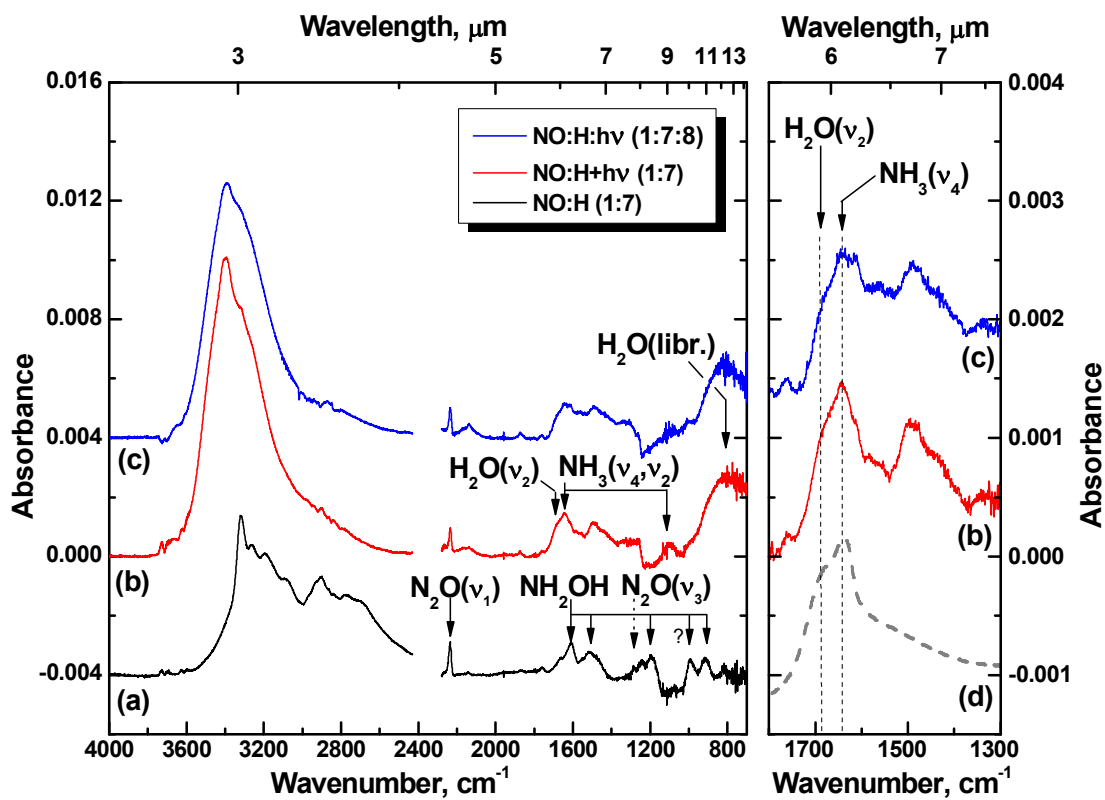

**Figure 1.** Left panel: RAIR spectra obtained after (a) co-deposition of NO molecules with H atoms (exp. 1.2); (b) co-deposition of NO molecules with H atoms followed by UV-photon exposure of $\sim$1.6x10$^{16}$ photons cm$^{-2}$ (exp. 1.3); (c) co-deposition of NO molecules and H atoms with simultaneous UV-photon exposure of $\sim$6.5x10$^{16}$ photons cm$^{-2}$ (exp. 1.4). Right panel: zoom-in of the spectral range 1800-1300 cm$^{-1}$ that shows the presence of newly formed $NH_3$ and $H_2O$ in the ice compared to a spectrum of a $NH_3:H_2O$ 1:4 mixture taken from Fig. 11.14 of Öberg 2009 (d). The total exposure dose of NO molecules is estimated to be 8.5 L in all three experiments. The total H-atom fluence is 6x10$^{16}$ atoms cm$^{-2}$.

Co-exposure of UV-irradiation and H atoms during deposition of NO molecules (Fig. 1c) results in a significant decrease in the amount of formed $NH_2OH$ (*i.e.*, only traces can be detected) and the formation of $H_2O$. Although $NH_3$ absorption features cannot be well distinguished in Fig. 1c, the corresponding TPD-QMS data indicate that the total area of m/z = 15 ($NH^+$), m/z = 16 ($NH_2^+$) or m/z = 17 ($NH_3^+$) signals of $NH_3$ desorbing between 100 and 140 K are nearly equal to those observed in the sequential exposure experiment; the values are within 10%. Furthermore the m/z = 15:16:17 signal ratio is equal to 0.06:0.84:1 at the desorption maximum. This is in good agreement with the values available from the NIST mass spectrometry data base[1], confirming the assignment of this desorption peak to $NH_3$. Similarly, $N_2$ TPD areas (m/z = 28) do not reveal any major difference between the two experiments. TPD results for m/z = 28 and m/z = 16 are presented in Fig. 2. Along with the desorption peak of $NH_3$ itself (between 100 and 140 K), the m/z =16 spectrum illustrates the desorption of two other species that can be clearly resolved by their characteristic desorption temperatures. These are $H_2O$ (a desorption feature between 140 and 160 K) and $NH_2OH$ (a desorption feature between 160 and 200 K). The latter becomes possible because of the higher detection sensitivity of TPD-QMS.

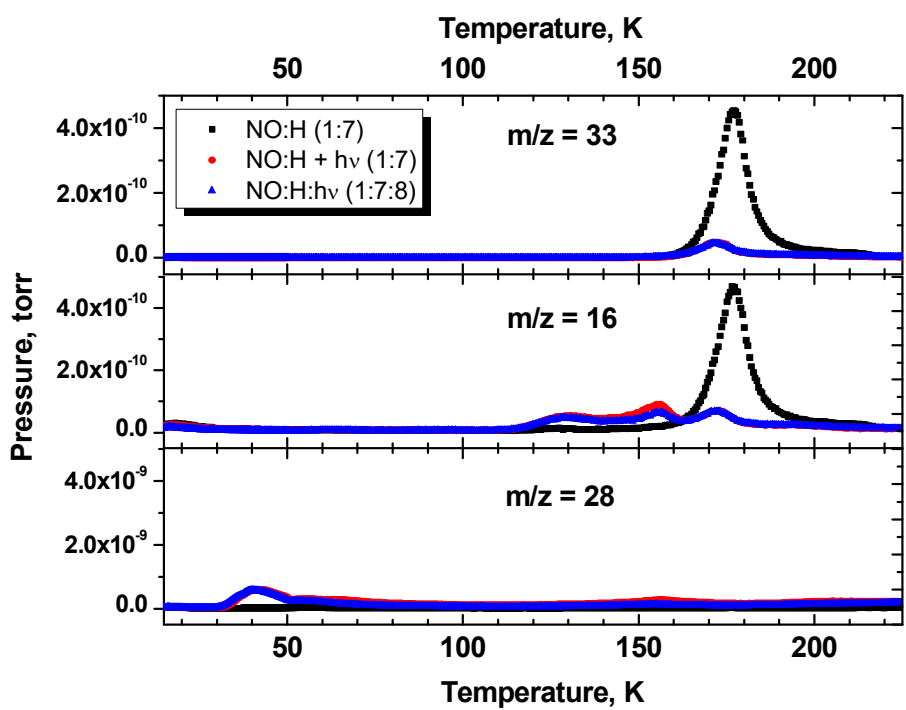

**Figure 2.** TPD spectra for three selected m/z values obtained after: *black squares* - co-deposition of NO molecules with H atoms (exp. 1.2); *red circles* - co-deposition of NO molecules with H atoms followed by UV-photon exposure of ~$1.6 \times 10^{16}$ photons $cm^{-2}$ (exp. 1.3); *blue triangles* - co-deposition of NO molecules and H atoms with simultaneous UV-photon exposure of ~$6.5 \times 10^{16}$ photons $cm^{-2}$ (exp. 1.4). The total exposure dose of NO molecules is estimated to be 8.5 L in all three experiments. The total H-atom fluence is $6 \times 10^{16}$ atoms $cm^{-2}$. The m/z = 33 signal corresponds to the $NH_2OH^+$ cation obtained by ionization of $NH_2OH$; the m/z = 16 signal corresponds to the $NH_2^+$ cation obtained by dissociative ionization of $NH_3$ and $NH_2OH$, or $O^+$ cation obtained by dissociative ionization of $H_2O$; the m/z = 28 signal corresponds to $N_2^+$.

Unfortunately, the RAIR data cannot be used to determine the $NH_2OH$ formation/destruction yield ratio in the experiments presented in Fig. 1 (exp. 1.2-1.4), because of the overlap of the hydroxylamine absorption features with those of water and ammonia. However, the same information can be obtained from QMS-TPD spectra (see Fig. 2).

---

[1]NIST Mass Spec Data Center, S. E. Stein, director, "Mass Spectra" in NIST Chemistry WebBook, NIST Standard Reference Database Number 69, Eds. P.J. Linstrom and W.G. Mallard, National Institute of Standards and Technology, Gaithersburg MD, 20899

The integration of the area of m/z = 33 signal in the range between 140 and 225 K (with a peak at 170 K) in both sequential and co-exposure experiments reveals that the final $NH_2OH$ yield is 10 times lower than in the case of co-deposition of NO and H atoms without UV irradiation.

**3.2. Simultaneous hydrogenation and UV processing of NO molecules co-deposited in a CO-rich environment**

Three sets of experiments are performed to investigate the co-exposure of UV photons and H atoms to NO molecules in CO ices. Experiments 2.1-2.3 presented in Fig. 3 are comparable to experiments 1.2-1.4 (shown in Fig. 1) with the exception that CO is added to the accreting gas mixture. Figure 3a shows a selected region of the acquired RAIR spectrum obtained after NO:CO = 1:12 gas-mixture deposition with H atoms. As for the RAIR spectrum in Fig. 1a, infrared absorption features of $NH_2OH$ in Fig. 3a can be clearly identified. The abundant presence of CO in the ice acts as a matrix, isolating the formed $NH_2OH$ molecules and preventing intermolecular (H-bond) interactions. Therefore, the $NH_2OH$ absorption bands are narrower and also somewhat shifted compared to those observed in Fig. 1a and, in accordance with Withnall & Andrews (1988) and Yeo & Ford (1990), these bands peak at 1606, 1375, 1142, and 889 $cm^{-1}$. As shown by Fedoseev et al. (2012), besides the fully saturated $NH_2OH$, also intermediate products of the hydrogenation scheme, such as HNO (1563 $cm^{-1}$), can be observed in the spectrum. This can be explained by the selected amount of co-deposited CO (12 times higher rate than NO) that may cause the trapping of intermediates like HNO in the bulk of the ice along with a suppression of ongoing hydrogenation.

In addition to the isolation effect of CO on NO and its hydrogenation products, CO can participate in H-atom addition reactions. Therefore, besides the strong absorption band of CO (centered at 2142 $cm^{-1}$) and its broad left shoulder presumably caused by lattice interactions, $H_2CO$ – a product of two H-atom additions to CO – can be successfully assigned through the absorption features centered at 1730 and 1496 $cm^{-1}$. As concluded in many previous studies (summarized in Hama & Watanabe 2013 and Linnartz et al. 2015) the CO hydrogenation results in the formation of methanol. In the present study methanol is not observed in the RAIRS data, but a QMS-TPD confirms its formation. As stated before, the presence of $N_2O$ in the spectrum is mainly caused by intrinsic contaminations in the NO gas bottle. No N-C bearing species are detected within the sensitivity range of our RAIR technique. Similar results were presented in Fedoseev et al. (2012).

The UV-irradiation of this ice (Fig. 3b) results in the disappearance of $NH_2OH$, HNO, and $H_2CO$ absorption features, consistent with their photodestruction, and the simultaneous appearance of new absorption features, originating from newly formed photoproducts. Besides the formation of HCO radicals and NO monomers, assigned by absorption bands centered at 1859 and 1091 $cm^{-1}$ (Ewing et al. 1960) and 1875 $cm^{-1}$ (Fateley et al. 1959, Minissale et al. 2014), respectively, the UV ice processing now results in the formation of N-C bearing molecules, such as HNCO. The latter is assigned through its strong absorption feature at 2261 $cm^{-1}$ (Teles et al. 1989). Additional spectroscopic information is added in Fig. 3 (right panels) to further constrain the assignments through comparison with spectra of HCO (obtained by UV-photoprocessing of $H_2$ entrapped in the solid CO matrix at similar experimental conditions) and NO monomers (Minissale et al. 2014). QMS data obtained from TPD experiments further support this assignment by detecting the HNCO desorption starting at 135 K and peaking at 160 K (see Fig. 4 for more details on the TPD-QMS data). The TPD-QMS spectrum also shows UV-photoproducts that cannot be identified by means of RAIRS. This is $N_2$ (note that the m/z = 14 signal is used instead of m/z = 28 that overlaps with the signal from CO) and a large desorption peak of $CO_2$ starting at around 68 K with its maximum peak at 81 K. Although $CO_2$ has strong absorption bands in the infrared, its assignment is challenged by the presence of absorption features from residual atmospheric $CO_2$ along the purged path of the infrared beam outside the main-chamber of the setup. The relative abundance of $CO_2$ in the irradiated ice sample, however, can be derived through the observation of $^{13}CO_2$ absorption at 2280 $cm^{-1}$ (see the corresponding zoom-in shown in Fig. 3). A natural isotope fractionation of roughly 1.1% with respect to regular $^{12}C$ is assumed. No clear signs of $NH_3$ and only traces of $NH_2OH$ are detected by means of QMS during TPD.

As opposed to the experiment involving solely NO (Fig. 1), the addition of CO molecules results in a different final ice composition in the case of the co-exposure experiment when compared to the sequential exposure experiment discussed above (see Fig. 3c and Fig. 3b for comparison). The simultaneous UV-photon and H-atom exposure of the co-deposited NO:CO gas mixture results in a significant reduction of the HNCO yield, as well as in a visual decrease of HCO and NO absorption features. The reduction of the HNCO yield is also confirmed by a decrease of the corresponding QMS integrated area (m/z = 43) obtained during TPD. Lower yields for HCO and NO can be expected. The co-exposure of the ice to UV photons and H atoms allows the species formed through the interaction with UV photons to react with H atoms. Thus the decrease in the formation yield of the species that are highly reactive towards H atoms is not surprising. HNCO, however, exhibits an activation barrier in the reaction with hydrogen atoms (Nguyen *et al.* 1996), so its decrease must be explained in a different way, *e.g.*, by the consumption of the intermediate species involved in its formation.

Integrating the area under the m/z = 33 ($NH_2OH$) signal in the range between 140 and 225 K for experiments 2.1-2.3, reveals a decrease of the $NH_2OH$ yield of about 10 times in the experiments 2.2 and 2.3 with respect to the yield measured in experiment 2.1. This is similar to what was found in experiments 1.2-1.4, *i.e.*, independent of the addition of CO.

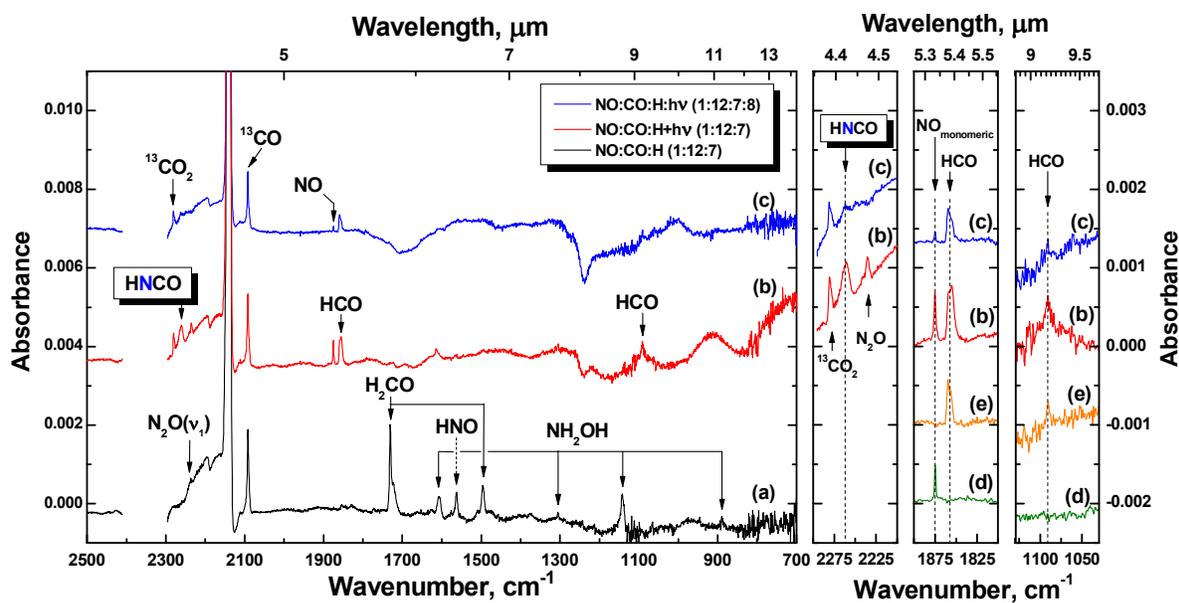

**Figure 3.** Left panel: RAIR spectra obtained after: (a) co-deposition of NO:CO = 1:12 molecular mixture with H atoms for the total ratio of 1:12:7 (exp. 2.1); (b) co-deposition of NO:CO = 1:12 molecular mixture with H atoms followed by the UV-photon exposure of ~1.6x10$^{16}$ photons cm$^{-2}$ (exp. 2.2); and (c) co-deposition of NO:CO = 1:12 molecular mixture and H atoms with simultaneous UV-photon exposure of ~6.5x10$^{16}$ photons cm$^{-2}$ (exp. 2.3). Right panels: zoom of the spectral ranges around 2250, 1850, and 1100 cm$^{-1}$ that show (from left to right) the presence of newly formed $^{13}CO_2$, HNCO, NO monomers and HCO through comparison with the spectrum of the isolated HCO obtained by UV-photoprocessing of $H_2$ entrapped in a solid CO matrix (e) and the spectrum of isolated NO taken from Minissale et al. (2014) (d). The total dose of NO molecules is estimated to be 8.5 L in all three experiments. The total estimated H-atom fluence is ~6x10$^{16}$ atoms cm$^{-2}$.

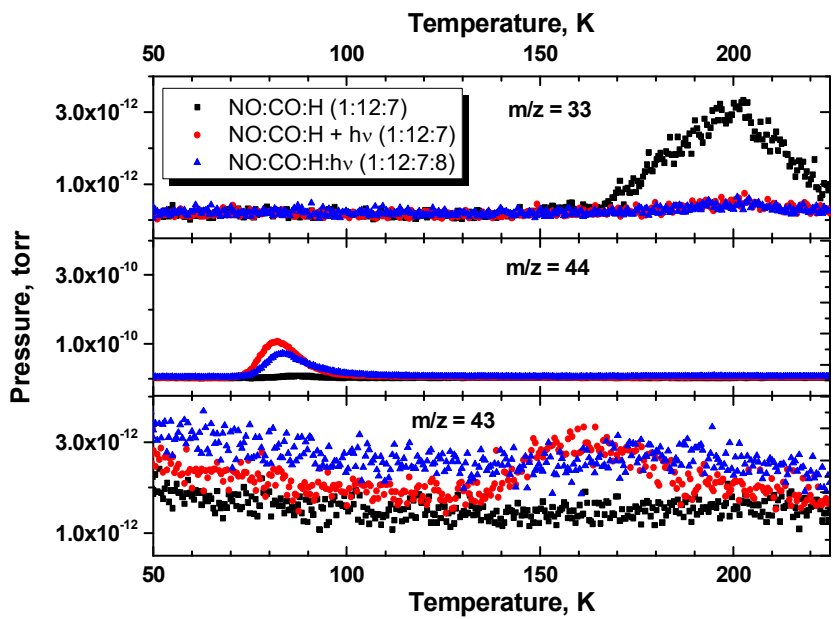

**Figure 4.** TPD spectra for three selected m/z values obtained after: *black squares* - co-deposition of NO:CO = 1:12 molecular mixture with H atoms for the total ratio of 1:12:7 (exp. 2.1) ; *red circles* - co-deposition of NO:CO = 1:12 molecular mixture with H atoms followed by the UV-photon exposure of ~$1.6 \times 10^{16}$ photons cm$^{-2}$ (exp. 2.2) ); *blue triangles* - co-deposition of NO:CO = 1:12 molecular mixture and H atoms with simultaneous UV-photon exposure of ~$6.5 \times 10^{16}$ photons cm$^{-2}$ (exp. 2.3). The total dose of NO molecules is estimated to be 8.5 L in all three experiments. The total H-atom fluence is $6 \times 10^{16}$ atoms cm$^{-2}$. Here m/z = 33; m/z = 44 and m/z = 43 correspond to $NH_2OH^+$, $CO_2^+$ and $HNCO^+$, respectively.

### 3.3. Simultaneous hydrogenation and UV processing of NO molecules co-deposited in H$_2$CO and CH$_3$OH-rich environments

Figure 5 presents the RAIR spectra obtained for experiments 3.2-3.3 (top panel) and 4.2-4.3 (bottom panel), where a CO environment (as in experiments 2.1-2.3) is replaced by H$_2$CO (3.2-3.3) and then by CH$_3$OH (4.2-4.3). In both the top and bottom panel, the results for the co-exposure experiments with and without UV irradiation are compared to each other. In Fig. 5a the RAIR spectrum obtained after co-deposition of an NO:H$_2$CO mixture with only H atoms shows the appearance of several new species. Along with the expected NH$_2$OH (see Fig. 1a for comparison) and H$_2$CO bands (1730, 1498, 1248, and 1178 cm$^{-1}$), absorption features of CH$_3$OH (see Öberg et al. 2009) can be successfully assigned in the spectrum. These are associated with the broad absorption in the range 1500-1400 cm$^{-1}$, caused by several ($\nu_6$, $\nu_4$, $\nu_5$, $\nu_{10}$) vibrational modes, and the strongest absorption of CH$_3$OH in the range between 1000 and 1070 cm$^{-1}$, caused by the C-O stretching vibration mode. It is not clear whether both of the absorption features in this spectral range are due to CH$_3$OH molecules in different environments or whether CH$_3$OH is responsible for only one of these features. Furthermore, carbon monoxide, an abstraction product of H-atom interaction with H$_2$CO, as found by Hidaka et al. 2004 and recently extensively discussed by Chuang et al. (2016), is also found (2136 cm$^{-1}$).

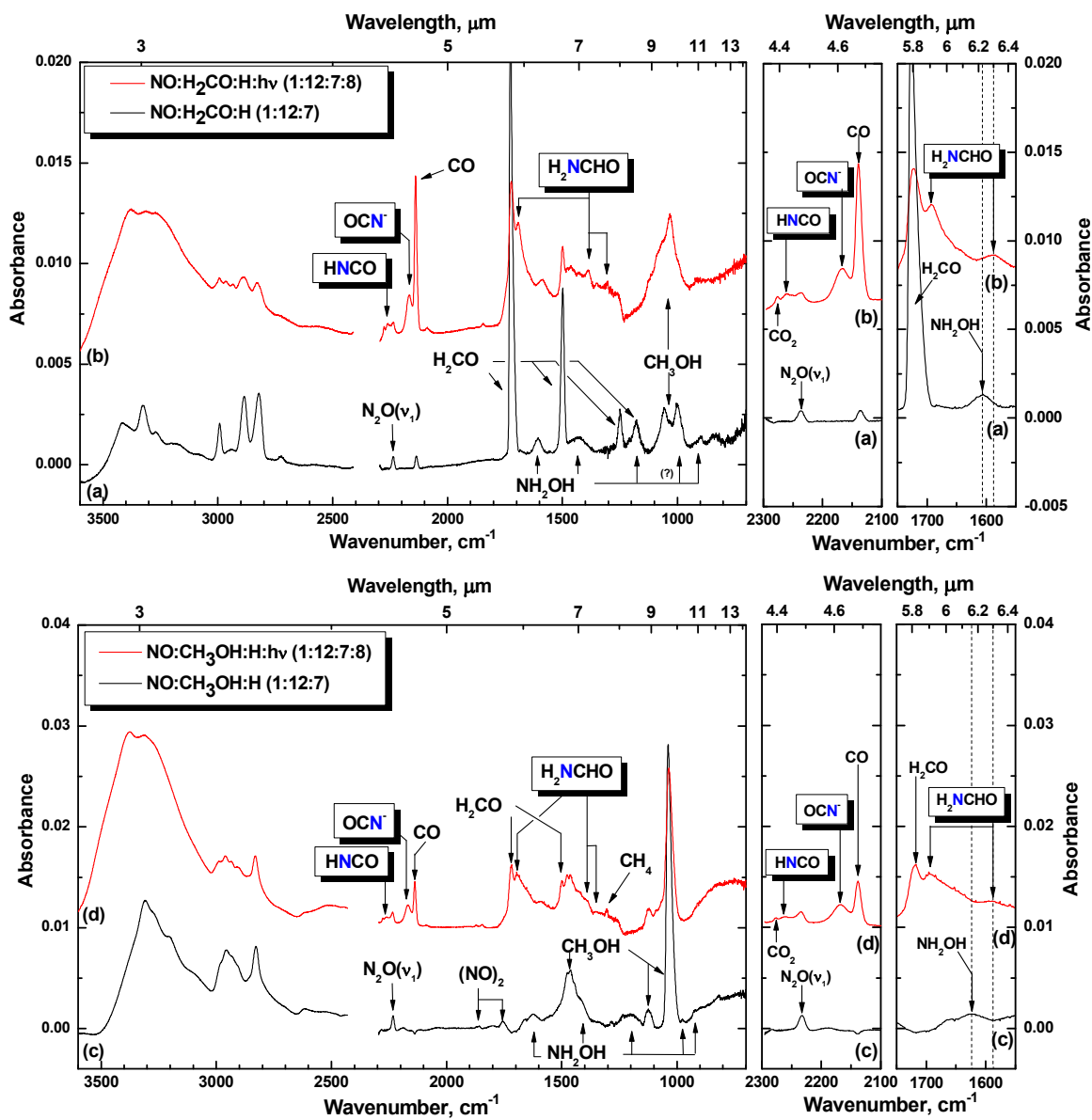

**Figure 5.** Top panel: RAIR spectra for NO in H₂CO obtained after: (a) co-deposition of NO:H₂CO = 1:12 molecular mixture with H atoms (exp. 3.2); (b) co-deposition of NO:H₂CO = 1:12 molecular mixture and H atoms with simultaneous UV-photon exposure (exp. 3.3). Bottom panel: RAIR spectra for NO in CH₃OH mixtures, obtained after: (c) co-deposition of NO:CH₃OH = 1:12 molecular mixture with H atoms (exp. 4.2); (d) co-deposition of NO:CH₃OH = 1:12 molecular mixture with simultaneous exposure of H atoms and UV photons (exp. 4.3). The total exposure dose of NO molecules is estimated to be 8.5 L in all the experiments. The total estimated H-atom and UV-photon fluences are ~6x10$^{16}$ atoms cm$^{-2}$ and ~6.5x10$^{16}$ photons cm$^{-2}$, respectively. The two right panels on top and bottom panel present a zoom-in of spectra in the 2200 and 1650 cm$^{-1}$ regions, where HNCO, OCN⁻, and H₂NCHO absorptions can be seen.

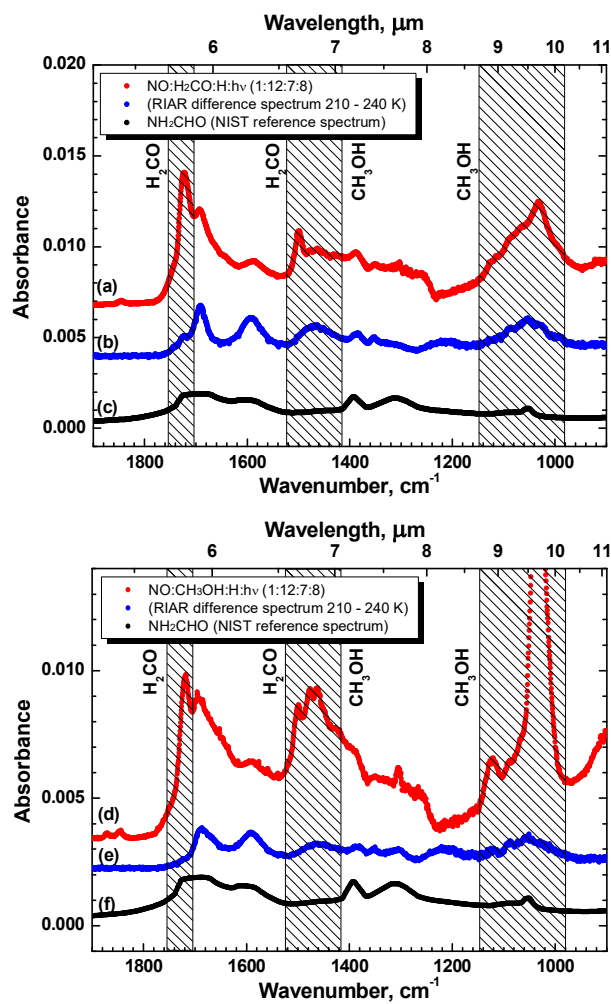

**Figure 6.** Top panel: RAIR spectra for (a) co-deposition of NO:H$_2$CO = 1:12 mixture with H atoms and UV photons (exp. 3.3); (b) RAIR difference spectra obtained between 210 and 240 K during the TPD after exp. 3.3; (c) NIST reference spectra of liquid NH$_2$CHO[2]. Bottom panel: (d) co-deposition of NO:CH$_3$OH = 1:12 mixture with H atoms and UV photons (exp. 4.3); (e) RAIR difference spectra obtained between 210 and 240 K during the TPD after exp. 4.3; (f) same as (c). Dashed areas indicate spectral regions where IR features from trapped H$_2$CO and CH$_3$OH at high temperatures are expected.

The simultaneous UV-processing of the ice (Fig. 5b) results in a significant decrease of H$_2$CO and NH$_2$OH absorption features and in the increase of CH$_3$OH and CO absorption bands. Simultaneously, new species are formed. These are N-C bearing species: HNCO (2261 cm$^{-1}$), which was already observed in the photolysis experiments involving CO (Figs. 3b-c); OCN$^-$ (2167 cm$^{-1}$, van Broekhuizen *et al.* 2004); and most importantly formamide, NH$_2$CHO (1693, 1590, 1388, and 1350 cm$^{-1}$, King *et al.* 1971, Sivaraman *et al.* 2013). Other irradiation products that can be assigned by means of RAIRS are CH$_4$ (1303 cm$^{-1}$) and $^{13}$CO$_2$ (2276 cm$^{-1}$). QMS data obtained during the TPD confirm a significant production of CO$_2$ during the photolysis as well. The CH$_4$ assignment is confirmed through QMS data by observing CH$_4$ desorption starting at 48 and peaking at 57 K with a m/z = 15 to m/z = 16 ratio of 0.9. Unfortunately, the broad and overlapping co-desorbing profiles of H$_2$CO, CH$_3$OH, and NH$_2$OH do not allow for a clear deconvolution of the HNCO and NH$_2$CHO desorption features. Furthermore, it should be noted that thermal processing of this ice results in a rich chemistry, since H$_2$CO exhibits strong chemical activity upon heating

---

[2] Evaluated Infrared Reference Spectra" in NIST Chemistry WebBook, NIST Standard Reference Database Number 69, Eds. P.J. Linstrom and W.G. Mallard, National Institute of Standards and Technology, Gaithersburg MD, 20899

in the presence of a base like $NH_3$ (see for example Schutte *et al.* 1993, Duvernay *et al.* 2014). In the present study $NH_2OH$ may play the role of interacting base. The most intense m/z signal of HNCO (m/z = 43) rises at 180 K and exhibits a maximum at 205 K (van Broekhuizen *et al.* 2004), while $NH_2CHO$ (m/z = 45) has at least two broad peaks in the TPD spectra with maxima at 165 and 215 K, confirming that both species may be present in the solid phase. The measured desorption peak is also in agreement with the desorption temperature of pure $NH_2CHO$, as reported by Dawley *et al.* (2014). On the other hand, RAIRS data show that formamide and isocyanic acid form efficiently already at 13 K, and well before thermal processing occurs during TPD.

The RAIR was obtained after NO hydrogenation in $CH_3OH$ ice (Fig. 5c) shows the presence of $CH_3OH$ (1037 $cm^{-1}$ and the broad absorption feature centered at 1465 $cm^{-1}$, see Öberg et al. 2009) and $NH_2OH$ (see Fig. 1a for comparison). Furthermore, non-hydrogenated NO in the form of *cis*-$(NO)_2$ can be assigned in the spectrum through absorptions at 1857 and 1756 $cm^{-1}$. The presence of NO indicates that HNO should also be found in the ice, however, its detection is not straight-forward because of the overlap between the strongest absorption band of HNO at 1562 $cm^{-1}$ and some $CH_3OH$ absorption features. The co-addition of UV photons in a similar experiment (Fig. 5d) results in the expected decrease of $CH_3OH$ and $NH_2OH$ absorption bands. The photoprocessing products are similar to those discussed in the case of $H_2CO$-rich ices. Besides the formation of $H_2CO$, CO, and $CH_4$ that may solely originate from $CH_3OH$ photoprocessing, three N-C bearing species are detected: $OCN^-$, HNCO, and $NH_2CHO$ (see Fig. 5b for comparison). It should be noted that the integrated areas under the absorption bands of $OCN^-$, HNCO, and $NH_2CHO$ are lower than in the case of $H_2CO$-rich ices. However, the final yields of these species are significantly higher than in the case of a CO-rich environment, where $NH_2CHO$ is not observed.

Similarly as in the case of $H_2CO$-rich ices, a broad desorption feature of $CH_3OH$ entrapping $H_2CO$ and $NH_2OH$ and eventual thermally-induced chemistry do not allow to confidently assign HNCO and $NH_2CHO$ desorbing peaks in the QMS data. Nevertheless, at least three peaks with m/z = 43 and m/z = 45 signals are present in the TPD spectra with desorption maxima at 135, 165, and 215 K, fully consistent with such an assignment. To further constrain the detection of $NH_2CHO$, the top panel of Fig. 6 compares a co-deposition of $NO:H_2CO$ = 1:12 mixture with H atoms and UV photons (exp. 3.3) with RAIR difference spectra obtained between 210 and 240 K during the TPD after exp. 3.3, and the NIST reference spectrum of liquid $NH_2CHO$. The bottom panel of Fig. 6 shows the analogue case for the $NO:CH_3OH$ = 1:12 mixture with H atoms and UV photons (exp. 4.3) compared to RAIR difference spectra obtained between 210 and 240 K during the TPD after exp. 4.3, and the NIST reference spectrum of liquid $NH_2CHO$. In both cases, the RAIR different spectra acquired at higher temperatures resemble closely the profile of the NIST spectrum of liquid $NH_2CHO$, confirming that $NH_2CHO$ is formed at low temperatures and that right before its desorption, a nearly pure $NH_2CHO$ ice is present on the gold substrate.

**4. Discussion**

The main products observed after UV-photolysis of hydrogenated NO ice, *i.e.*, $N_2$, $NH_3$, and $H_2O$, are similar to the main products observed during $NH_2OH$ gas-phase photolysis (Betts & Back 1965). Furthermore, no major differences are found between sequential and co-exposure experiments in terms of photoproducts of $NH_2OH$ ice. Therefore, it is concluded that the UV photons dissociate $NH_2OH$ and the resulting radicals can react with other ice components to form new species, while the UV photons impinging on NO and its unsaturated hydrogenation products do not result in the formation/detection of new molecules. Betts & Back (1965) investigated a series of high-pressure gas-phase reactions that may have an equivalent in the solid phase and could be relevant to the experiments discussed here. In their work, Betts & Back (1965) showed that the photodissociation of $NH_2OH$ leads mainly to the formation of $NH_2$ and OH radicals. These radicals are then involved in further reactions:

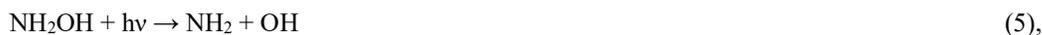

$$NH_2OH + h\nu \rightarrow NH_2 + OH \qquad (5),$$

$$NH_2 + NH_2OH \rightarrow NH_3 + NHOH \qquad (6),$$

$$OH + NH_2OH \rightarrow H_2O + NHOH \qquad (7),$$

$$NHOH + NHOH \rightarrow N_2 + 2H_2O \qquad (8).$$

Our experimental results do not contradict these routes (reactions 5-8). It should be noted that the energy of the N-O bond in NH$_2$OH (2.7 eV, Hettama & Yarkony 1995) is significantly lower than the energy of the UV photons produced by our H$_2$-discharge lamp (10.2 and ~7.7 eV, Ligterink et al. 2015b). Therefore, if photodissociation of NH$_2$OH proceeds through reaction (5), one or even both of the formed NH$_2$ and OH radicals could be in their electronically and vibrationally excited states (Gericke et al. 1994, Luckhaus et al. 1999) or have excess translational energy. This, in turn, may promote reactions (6) and (7) that both require activation barriers to proceed in the solid phase. Moreover, the absence of a significant HNO and NO yield indicates that the photodissociation of NH$_2$OH into HNO and two H atoms (Gericke et al. 1994) is an overall less efficient pathway:

$$NH_2OH + h\nu \rightarrow HNO + 2H \qquad (9),$$

$$HNO + h\nu \rightarrow NO + H \qquad (10).$$

The photodissociation of NH$_2$OH into NH$_2$ and OH (reaction 5) finds further confirmation when CO molecules are added into the chemical network. Both HNCO and CO$_2$ molecules observed in the experiments 2.2-2.3 can be formed by the interaction of NH$_2$ and OH radicals with the abundantly present CO molecules:

$$OH + CO \rightarrow CO_2 + H \qquad (11),$$

$$NH_2 + CO \rightarrow HNCO + H \qquad (12).$$

The formation of CO$_2$ as a result of CO interaction with OH is reported in a number of experimental studies and is expected to proceed through the HOCO complex as an intermediate that has a low activation barrier and can therefore be formed through reaction (11) with OH in its ground electronic state (Goumans et al. 2008, Oba et al. 2010, Ioppolo et al. 2011, Noble et al. 2011). Recent molecular dynamics calculations (Arasa et al. 2013) showed that the spontaneous dissociation of the HOCO complex requires an activation barrier to occur. Thus, in the formation of CO$_2$, a two-step mechanism may be involved in which a HOCO complex is formed by interaction of CO with a hydroxyl radical followed by an interaction with H-atoms or UV-photons. Calculations show that reaction (12) is endothermic by 0.41 eV and that it has an activation barrier of 1.39 eV (Nguyen et al. 1996). Thus it requires NH$_2$ to be in a vibrationally or electronically excited state (or NH$_2$ with a very high translational energy) as will be discussed below, and reaction (12) to proceed before deexcitation of NH$_2$ occurs. The consumption of OH and NH$_2$ in reactions (11) and (12) also explains the decrease in the NH$_3$, H$_2$O, and consequently N$_2$ final yields as observed in the photolysis experiments involving CO, compared to the values found in the corresponding photolysis experiments of pure NO hydrogenation products.

The addition of H$_2$CO or CH$_3$OH into the photolysis scheme complicates the reaction network because H$_2$CO and CH$_3$OH are also easily photodissociated resulting in efficient formation of several other photoproducts. The two possible H$_2$CO photodissociation channels are:

$$H_2CO + h\nu \rightarrow HCO + H \qquad (13),$$

$$H_2CO + h\nu \rightarrow CO + H_2 \quad (14),$$

(see McQuigg & Calvert 1969, Moore & Weisshaar 1983, Rhee *et al.* 2007). HCO and CO, both $H_2CO$ photodissociation products, are chemically active towards $NH_2$ and OH. Interactions with CO result in the formation of HNCO and $CO_2$ following reactions (11) and (12), as discussed earlier in this section. HNCO and $CO_2$ are indeed among the products observed during the simultaneous photolysis and hydrogenation of NO in $H_2CO$-rich ice.

Formamide is another product observed in this study that may be formed through the interaction of $NH_2$ and HCO:

$$NH_2 + HCO \rightarrow H_2NCHO \quad (15a).$$

This radical-radical recombination should proceed without any activation barrier, however, a competing reaction should not be neglected:

$$NH_2 + HCO \rightarrow NH_3 + CO \quad (15b).$$

Similarly to $NH_2$, the interaction of OH with HCO may proceed through two different pathways:

$$OH + HCO \rightarrow HC(O)OH \quad (16a),$$

$$OH + HCO \rightarrow H_2O + CO \quad (16b).$$

Although no HCOOH and $NH_3$ detections are possible in this study (see section 3.3), their formation cannot be disproven due to the overlap of typical RAIR absorption features with those of other more abundant molecules. Thus no solid conclusion can be made on the efficiency and branching ratio of reactions (15)-(16).

$NH_2$ and OH radicals formed in reaction (5) should be produced with significant excess energy. This excess energy could contribute to overcoming of activation barriers for the reactions of $NH_2$ and OH with $H_2CO$. In the first case, $NH_2CHO$ should be formed (Barone *et al.* 2015):

$$NH_2 + H_2CO \rightarrow H_2NCHO + H \quad (17a).$$

However, similar to reactions (15) and (16), a competing abstraction reaction may occur:

$$NH_2 + H_2CO \rightarrow NH_3 + HCO \quad (17b).$$

The interaction of OH with $H_2CO$ likely results in the abstraction reaction (Hudson *et al.* 2006):

$$OH + H_2CO \rightarrow H_2O + HCO \quad (18).$$

Recent calculations by Barone *et al.* (2015) suggested that reaction (17a) is nearly barrierless and can be efficient even at very low temperatures. If this is indeed the case, formamide production *via* reaction (5) followed by reaction (17a) should be overall more efficient than through a pathway involving reactions (5), (13) and (15a).

$OCN^-$ observed in this study is a direct derivate of HNCO and can be formed in a number of proton transfer reactions. Along with $H_2O$ and $NH_3$ that both can act as a base in interaction with isocyanic acid (Grim *et al.* 1989, Mispilier *et al.* 2012, Theule *et al.* 2011), $NH_2OH$ and $NH_2CHO$ can act as a base as well:

$$\text{NH}_2\text{OH (NH}_3, \text{H}_2\text{O, NH}_2\text{CHO)} + \text{HNCO} \rightarrow \text{NH}_3\text{OH}^+ (\text{NH}_3^+, \text{H}_3\text{O}^+, \text{NH}_3^+\text{CHO)} \text{NCO}^- \qquad (19).$$

This can explain the detection of OCN$^-$ in our RAIR data. The excess energy of HNCO produced through reaction (12) or through other possible mechanisms should help to overcome the activation barrier of reaction (19). A ratio between HNCO and OCN$^-$ transmission band strengths of 1.7 is known from the literature (van Broekhuizen et al. 2005) and indicates that most of the isocyanic acid is in the form of its anion (see Figs. 5b and 5d).

Simultaneous photolysis and hydrogenation of NO in a CH$_3$OH environment results in the same products observed in the two previous discussed cases, *i.e.*, the photolysis and hydrogenation of NO in a CO and H$_2$CO-rich environment. This is not surprising because photoprocessing of CH$_3$OH leads to the dehydrogenation products H$_2$CO and CO and thus to the two cases already discussed before. The detection of methane among the photolysis products shows that part of the methanol is also photodissociated into OH and CH$_3$ to yield CH$_4$. Moreover, the interaction of CH$_3$OH with the photoproducts of reaction (5), *i.e.*, energetically excited NH$_2$ and OH, should lead to H-atom abstraction reactions from methanol, due to the saturated nature of CH$_3$OH molecules. The lower formation yields of HNCO, OCN$^-$, and NH$_2$CHO in comparison to the case of H$_2$CO-rich ice is in agreement with the picture that the interaction of NH$_2$ and OH with H$_2$CO and CO is responsible for the observed reaction products rather than the direct interaction of NH$_2$ with CH$_3$OH molecules. Similar to the sequential and co-exposure experiments in H$_2$CO-rich ices, most of the HNCO ends up in its anionic form, OCN$^-$.

To conclude this discussion section we mention another aspect of CH$_3$OH-related photochemistry. The photodissociation of CH$_3$OH also can lead to the formation of three distinct carbon-bearing radicals, *i.e.*, CH$_3$, CH$_2$OH, and CH$_3$O. The formation of CH$_3$ finds direct confirmation through the detection of CH$_4$ that is formed in the reaction:

$$\text{CH}_3 + \text{H} \rightarrow \text{CH}_4 \qquad (20),$$

or through the abstraction of H-atoms by CH$_3$ radicals from the other species present in the ice. Interactions of CH$_3$, CH$_2$OH, and CH$_3$O with NH$_2$ radicals, formed in the reaction (5), may in turn yield methylamine:

$$\text{CH}_3 + \text{NH}_2 \rightarrow \text{CH}_3\text{NH}_2 \qquad (21),$$

and two other rather exotic species. These are aminomethanol:

$$\text{CH}_2\text{OH} + \text{NH}_2 \rightarrow \text{NH}_2\text{CH}_2\text{OH} \qquad (22),$$

and methoxyamine:

$$\text{CH}_3\text{O} + \text{NH}_2 \rightarrow \text{CH}_3\text{ONH}_2 \qquad (23).$$

Unfortunately, none of these complex molecules can be confidently detected in our RAIR spectra at 13 K. There is a significant overlap between the main infrared absorption features of CH$_3$NH$_2$ and CH$_3$ONH$_2$ (Nelson 1970, Gray & Lord 1957) with infrared bands of other species clearly present in the ice. As far as we know, IR spectra are not available in the literature for H$_2$NCH$_2$OH obtained by direct deposition due to the high instability of this species at room temperature. Nevertheless, based on other work, predictions can be made. Recent work by Paardekooper et al. (2016) report effective branching ratios of CH$_3$OH photodissociation with CH$_2$OH:CH$_3$:CH$_3$O equal to 8:3:2 for UV-sources similar to that used in the present study. This makes H$_2$NCH$_2$OH (reaction 22) and to a lesser extent CH$_3$NH$_2$ (reaction 21) other likely candidates to be formed in experiments involving simultaneous hydrogenation and photolysis of NO in a CH$_3$OH environment.

**5. Astrochemical implication and conclusions**

Various astrochemical models show that a maximum in the gas-phase NO abundance is reached during the late stages of dark molecular cloud evolution (Charnley *et al.* 2001, Congiu *et al.* 2012b, Vasyunina *et al.* 2012), when the so-called "CO freeze-out" occurs (Tielens *et al.* 1991, Gibb *et al.* 2004, Pontoppidan 2006, Öberg *et al.* 2011, Visser *et al.* 2011, Boogert *et al.* 2015). Thus a peak of NO accretion on the grain surface is expected simultaneously with CO accretion and the chemical environment around depositing NO molecules should be rich in molecules observed in the CO-rich layer of the ice mantle. $CH_3OH$ and $H_2CO$, *i.e.*, the main hydrogenation surface reaction products of solid CO, are also expected to be present in the same interstellar ice layer (Cuppen *et al.* 2011, Penteado *et al.* 2015).

Here we show that the accretion of NO molecules under physical-chemical conditions that are similar to those observed in dark molecular clouds (*i.e.*, $T \leq 15$ K, simultaneous H-atom accretion and UV photon exposure, ice layers rich in CO, $H_2CO$, and $CH_3OH$ molecules) results in the efficient formation of three N-C bearing species: HNCO, $OCN^-$, and $NH_2CHO$. $OCN^-$ is usually assigned to the full or at least part of the XCN band that is observed in interstellar ices towards both low- and high-mass young stellar objects with abundances ranging from 0.3 to 0.8% with respect to water ice (Öberg *et al.* 2011, Boogert *et al.* 2015). Although HNCO has not been observed in the solid phase, it is referred to as a direct precursor for $OCN^-$ (van Broekhuizen *et al.* 2005, Boogert *et al.* 2015). The higher abundance of $OCN^-$ over HNCO in our experiments is consistent with astronomical observations. Another important outcome of our study is the formation of $NH_2CHO$. If accretion of NO molecules onto grain surfaces is partially responsible for the $OCN^-$ observations in ice mantles, then our work shows that $NH_2CHO$ should also be present as a product of combined NO hydrogenation and photolysis in CO-rich ices and therefore linked to the formation of HNCO and $OCN^-$. Correlations between HNCO:$OCN^-$:$NH_2CHO$ abundances in the solid state will result in the observation of HNCO:$NH_2CHO$ ratio in the gas phase upon sublimation of the ice. This result is fully in line with the observed gas-phase correlation of HNCO and $NH_2CHO$ abundancies in the protostellar shock regions L1157-B1 and L1157-B2 (Lopez-Sepulcre *et al.* 2015) as well as in low- and high-mass pre-stellar and protostellar objects (Bisschop *et al.* 2008, Mendoza *et al.* 2014). Moreover, it should be noted that although solid-state detections of HNCO and $NH_2CHO$ in interstellar ices are lacking and only upper limits have been derived (Schutte *et al.* 1999, van Broekhuizen *et al.* 2004), both of these species have been observed in cometary ices with abundances of about 0.02 and 0.002 with respect to water ice (Mumma & Charnley 2011, Goesman *et al.* 2015, le Roy *et al.* 2015).

Figure 7 summarizes all findings in a reaction network illustrating how the aforementioned N-C bearing complex molecules HNCO, $OCN^-$ and $NH_2CHO$ can be formed. The central role $NH_2$ plays is clear from this figure. In the experiments described here it is formed through the photodissociation of hydroxylamine, formed upon NO hydrogenation, but in interstellar clouds $NH_2$ radical could also originate from the photodissociation of $NH_3$, or from hydrogenation of atomic N. In the latter case, the $NH_2$ would not be energetic, reducing the probability to react with CO to form HNCO. Laboratory experiments showed that non-energetic reaction routes to HNCO and $NH_2CHO$ are NH + CO, $NH_2$ + HCO and, possibly, $NH_2$ + $H_2CO$ (see discussion here and Fedoseev *et al.* 2015). However, as interstellar $NH_3$ will be mostly mixed with $H_2O$ in water-rich layers, and N hydrogenation is largely expected to occur in such polar ices (i.e., CO-poor ices), the present scheme, starting from NO, remains the most likely one that shows a chemical link among HNCO, $OCN^-$, and $NH_2CHO$.

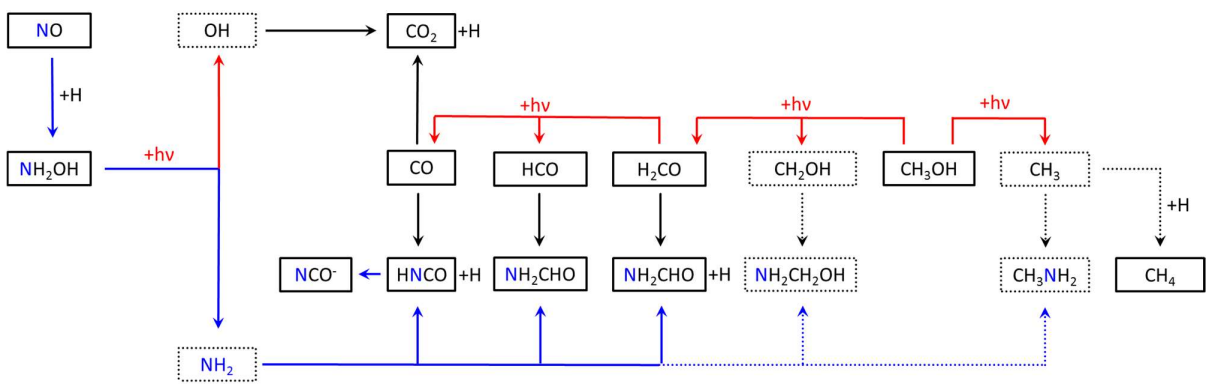

**Figure 7.** Proposed solid state reaction network obtained from this study showing how various N-C bearing complex organic molecules are formed starting from hydrogenation and UV irradiation of NO ice. Solid boxes mark molecules and free radicals detected in this study, while dotted boxes indicate species for which direct laboratory detections are still missing. Blue arrows indicate nitrogen-bearing species formation routes. Red arrows indicate photodissociation routes.

The formation of two other complex organic species - aminomethanol ($NH_2CH_2OH$) and methylamine ($CH_3NH_2$) – is indicated in Fig. 7 as well. The intermediate free radicals required to form these species are present in the ice. Nevertheless, the detection of aminomethanol and methylamine in our experiments has not been successful yet.

Besides aforementioned formation of N-C bearing complex molecules, co-exposure of NO molecules to UV photons and H atoms results in a significant decrease of $NH_2OH$ in the ice. In particular, a drop of one order of magnitude in $NH_2OH$ abundance is observed under our experimental conditions, when $NH_2OH$ is exposed to UV photons. This also puts an upper limit on the amount of accreted NO molecules that can be converted into HNCO, $OCN^-$ and $NH_2CHO$; this value has to be less or equal than the loss in $NH_2OH$ abundance. Moreover, the efficient photodissociation of $NH_2OH$ and the observed reactivity of its photofragments in CO, $H_2CO$, and $CH_3OH$-rich ices may account for a lower $NH_2OH$ abundance in interstellar ices than expected on the basis of hydrogenation experiments only (Congiu *et al* 2012a) and can be one of the reasons responsible for the so far unsuccessful efforts to identify this molecule in the ISM (Pulliam *et al.* 2012, McGuire *et al.* 2015).

The last point that is stressed here is the astrobiological importance of our findings. The general role that molecules containing a N-C bond could play in the synthesis of amino acids and nucleobases is well illustrated by the example of HCN polymerization (see for instance Figure 1 of Matthews & Minard 2008). In turn, a [-(H)NC(O)-] functional group is a necessary constituent of all peptides – oligomers and polymers consisting of amino-acids that are crucial for life on Earth. Thus the formation of molecules with an [-(H)NC(O)-] structural backbone, such as HNCO and $NH_2CHO$, is interesting from an astrobiological point of view because these molecules are potentially involved in the synthesis of fragments of peptides as schematically illustrated in Figure 6 of Fedoseev *et al.* (2015). Clearly, the detection of interstellar ice constituents with peptide (-(H)NC(O)-) or just N-C bonds with the James Webb Space Telescope (JWST) would provide new insights in chemical processes that are considered to provide the building blocks of biologically relevant molecules. Also in the laboratory, ices rich in HNCO ($OCN^-$), $NH_2CHO$, and $CH_3OH$ offer a starting point for studies that focus on the formation of 'heavier' species. Such work reaches beyond the potential importance of interstellar ices, as HNCO and $NH_2CHO$ delivered to primordial terrestrial oceans through cometary impacts may also play a role in cometary ices, full in line with recent studies analyzing the constituents of comet 67P/Churyumov–Gerasimenko (Goesman *et al.* 2015, le Roy *et al.* 2015).

**Acknowledgments**

We express our gratitude for financial support through a VICI grant of NWO (the Netherlands Organization for Scientific Research), by NOVA (the Netherlands Research School for Astronomy), A-ERC grant 291141 "CHEMPLAN" and the FP7 ITN "LASSIE" (GA 238258). SI acknowledges the Royal Society and the STSM COST action CM1401 for financial support.